%% file: main.tex
\RequirePackage{amsmath}
\documentclass[runningheads,11pt]{llncs}

\input{header}

\begin{document}

\title{No Questions Asked: Effects of Transparency on Stablecoin Liquidity During the Collapse of Silicon Valley Bank}

\titlerunning{No Questions Asked: Effects of Transparency on Stablecoin Liquidity}

\input{author}

\maketitle

\input{abstract}

\input{sections/introduction}

\input{sections/related_work}

\input{sections/background}

\input{sections/methodology}

\input{sections/results}

\input{sections/discussion}

\input{sections/limitations}

\input{sections/conclusion}

\bibliographystyle{splncs04}
\bibliography{references}

\end{document}

%% file: header.tex
\usepackage[T1]{fontenc}

\usepackage{multirow}
\usepackage{graphicx}
\usepackage{geometry}
\usepackage{booktabs}
\usepackage{subcaption}
\usepackage{csquotes} %

\usepackage{acro}[=v2]

\usepackage{hyperref}

\usepackage{amsmath}

\usepackage[flushleft]{threeparttable}

\DeclareAcronym{US}{
    short = US,
    long = United States
}

\DeclareAcronym{fed}{
    short = FED,
    long = Federal Reserve
}

\DeclareAcronym{fomc}{
    short = FOMC,
    long = Federal Reserve Open Market Committee
}

\DeclareAcronym{usd}{
    short = USD,
    long = United States Dollar
}

\DeclareAcronym{mmf}{
    short = MMF,
    long = Money Market Fund
}

\DeclareAcronym{cnav}{
    short = CNAV,
    long = Constant Net Asset Value
}

\DeclareAcronym{nav}{
    short = NAV,
    long = Net Asset Value
}

\DeclareAcronym{svb}{
    short = SVB,
    long = Silicon Valley Bank
}

\DeclareAcronym{tether}{
    short = Tether,
    long = Tether Holdings Limited
}

\DeclareAcronym{fcb}{
    short = FCB,
    long = First Citizens Bank
}

\DeclareAcronym{fdic}{
    short = FDIC,
    long = Federal Deposit Insurance Corporation
}

\DeclareAcronym{dlt}{
    short = DLT,
    long = Distributed Ledger Technology
}

\DeclareAcronym{dlts}{
    short = DLTs,
    long = Distributed Ledger Technologies
}

\DeclareAcronym{tradfi}{
    short = TradFi,
    long = Traditional Finance
}

\DeclareAcronym{defi}{
    short = DeFi,
    long = Decentralized Finance
}

\DeclareAcronym{amm}{
    short = AMM,
    long = Automated Market Maker
}

\DeclareAcronym{dexs}{
    short = DEXs,
    long = Decentralized Exchanges
}

\DeclareAcronym{dex}{
    short = DEX,
    long = Decentralized Exchange
}

\DeclareAcronym{tvl}{
    short = TVL,
    long = Total Value Locked
}

\DeclareAcronym{cexs}{
    short = CEXs,
    long = Centralized Exchanges
}

\DeclareAcronym{lob}{
    short = LOB,
    long = Limit Order Book
}

\DeclareAcronym{lobs}{
    short = LOBs,
    long = Limit Order Books
}

\DeclareAcronym{nft}{
    short = NFT,
    long = Non-Fungible Token
}

\DeclareAcronym{nfts}{
    short = NFTs,
    long = Non-Fungible Tokens
}

\DeclareAcronym{icos}{
    short = ICOs,
    long = Initial Coin Offerings
}

\DeclareAcronym{usdc}{
    short = USDC,
    long = USD Coin
}

\DeclareAcronym{usdt}{
    short = USDT,
    long = Tether USDT
}

\DeclareAcronym{dai}{
    short = DAI,
    long = Dai Stablecoin
}

\DeclareAcronym{busd}{
    short = BUSD,
    long = Binance USD
}

\DeclareAcronym{tusd}{
    short = TUSD,
    long = TrueUSD
}

\DeclareAcronym{wbtc}{
    short = WBTC,
    long = Wrapped Bitcoin
}

\DeclareAcronym{weth}{
    short = WETH,
    long = Wrapped Ethereum
}

\DeclareAcronym{ai}{
    short = AI,
    long = Artificial Intelligence
}

\DeclareAcronym{nlp}{
    short = NLP,
    long = Natural Language Processing
}

\DeclareAcronym{did}{
    short = DiD,
    long = Difference-in-Differences
}

\DeclareAcronym{mci}{
    short = MCI,
    long = Marginal Cost of Immediacy
}

\DeclareAcronym{mmmfs}{
    short = MMMFs,
    long = Mutual Market Funds
}

\DeclareAcronym{mmmf}{
    short = MMMF,
    long = Mutual Market Fund
}

\DeclareAcronym{us}{
    short = US,
    long = United States
}

\DeclareAcronym{rv}{
    short = RV,
    long = Robustness Value
}

\DeclareAcronym{circle}{
    short = Circle,
    long = Circle Internet Financial Ltd.
}

%% file: author.tex
\author{Walter Hernandez Cruz\inst{1,2}
\and
Jiahua Xu\inst{1,2}
\and
Paolo Tasca\inst{1,2}
\and
Carlo Campajola\inst{1,2}
}

\authorrunning{Hernandez Cruz, Campajola et al.}

\institute{University College London, London, UK \and
DLT Science Foundation, London, UK\\
\email{\{walter.hernandez.18, jiahua.xu, p.tasca, c.campajola\}@ucl.ac.uk}}

%% file: abstract.tex
\begin{abstract}

Fiat-pegged stablecoins are by nature exposed to spillover effects during market turmoil in \ac{tradfi}.
We observe a difference in \ac{tradfi} market shocks impact between various stablecoins, in particular, \ac{usdc} and \ac{usdt}, the former with a higher reporting frequency and transparency than the latter.
We investigate this, using top \ac{usdc} and \ac{usdt} liquidity pools in Uniswap, by adapting the Marginal Cost of Immediacy (MCI) measure to Uniswap's Automated Market Maker, and then conducting Difference-in-Differences analysis on MCI and \ac{tvl} in \acs{usd}, as well as measuring liquidity concentration across different providers.
Results show that the \ac{svb} event reduced \ac{usdc}'s \ac{tvl} dominance over \ac{usdt}, increased \ac{usdt}'s liquidity cost relative to \ac{usdc}, and liquidity provision remained concentrated with pool-specific trends.
These findings reveal a flight-to-safety behavior and counterintuitive effects of stablecoin transparency: \ac{usdc}'s frequent and detailed disclosures led to swift market reactions, while \ac{usdt}'s opacity and less frequent reporting provided a safety net against immediate impacts.

\keywords{Stablecoins \and Financial Contagion \and Liquidity Risk \and Decentralized Exchanges \and Investor Behavior \and Decentralized Finance \and Market Microstructure.}

\end{abstract}

%% file: sections/introduction.tex
In March 2023, \ac{svb}, the leading commercial bank servicing nearly half of all venture-backed tech startups in Silicon Valley, collapsed \cite{Tobin2023SiliconBloomberg}. During \ac{svb}'s collapse \ac{circle}, the issuer of the second-largest stablecoin by market capitalization \ac{usdc} \cite{TopStablecoinsMarketCap}, revealed that nearly 8\% of its cash reserves \cite{USDCReserveMarch2023,Capoot2023StablecoinExposure} amounting to US\$3.3 billion was held at \ac{svb} and had been frozen - and potentially lost as uninsured deposits\footnote{https://twitter.com/circle/status/1634391505988206592}.

Stablecoins are digital assets (\blockquote{tokens}) used in cryptocurrency markets as proxies for fiat money. This is a necessity since, for both regulatory and technological reasons, it is impossible to use fiat money for operations on blockchains. Their value is pegged to a currency, typically the US Dollar, and the peg is maintained by backing the asset with reserves. Usually, these can be in fiat money (\blockquote{fiat-backed} stablecoins), or cryptocurrencies like Ether (\blockquote{crypto-backed} stablecoins), and stablecoin holders have the right to redeem their tokens for the equivalent underlying upon request in a structure that is similar to that of a \ac{cnav} \ac{mmf}. The most popular fiat-backed stablecoins at the time of writing are \ac{tether}'s \ac{usdt} and Circle's \ac{usdc}, while MakerDAO's \ac{dai} is the most used in the crypto-backed family. Much like the runs that \acs{cnav} \acs{mmf}s suffer when their \ac{nav} \blockquote{breaks the buck} \cite{Schmidt2016RunsFunds}, \ac{circle}'s transparency in declaring their significant exposure as an uninsured depositor of \ac{svb} led to a panic in cryptocurrency markets, causing \ac{usdc} to lose its peg to the US Dollar and trade below US\$0.87 for several hours \cite{Bonifacic2023USDCCollapse}.

Our study analyzes the impact that \ac{svb}'s collapse had on liquidity provision in \ac{dexs}. In particular, we analyze the highest-volume liquidity pools trading the two most popular stablecoins, \ac{usdc} and \ac{usdt}, and compare the dynamics of concentration and depth of available liquidity in the weeks leading to and following the event. Liquidity pools are trading venues that operate through an \ac{amm}, a set of algorithms matching liquidity providers with liquidity takers without relying on a centralized market maker or clearinghouse \cite{Xu2023SoK:Protocols,Dahi2023AutomatedLedger}. The most popular of these \ac{dexs} is Uniswap v3 \cite{Adams2021UniswapCore}, which at the time of the event dominated the market with over 60\% of \ac{dexs} trading volume \cite{Lee2023MarketCoinGecko} and is our main source of liquidity data.

Our study finds its motivation in several streams of financial economics literature. In particular, we are interested in analyzing the impact of information asymmetries \cite{Akerlof1970TheMechanism,Holmstrom2019UnderstandingSystem,DIAMOND1991DisclosureCapital} on liquidity provision, a topic that has been extensively studied in the context of traditional, \ac{lob}-based markets \cite{Campajola2020UnveilingNetworks,Drechsler2020LiquidityVolatility} but, to the best of our knowledge, has so far remained largely unexplored in \ac{defi} markets. We also take inspiration from the literature on financial contagion \cite{Allen2000FinancialContagion,Brunnermeier2009MarketLiquidity}, as we analyze the spillovers of liquidity shocks from the real world to cryptocurrency markets, and from the extensive literature on market microstructure that focuses on optimal execution and liquidity dynamics \cite{Amihud1986AssetSpread,Vayanos2012LiquidityCompetition}. Throughout our analysis, we consider \ac{usdc} as our asset of interest and keep \ac{usdt} as a control. This choice is dictated by the fact that the two assets are similar in most aspects (i.e. type of backing, adoption, liquidity) except for the exposure that \ac{usdc} disclosed to the \ac{svb} bankruptcy\cite{USDCReserveMarch2023,Capoot2023StablecoinExposure}. Noticeably, it is unknown whether or not \ac{usdt} was also exposed to the event\cite{USDTReserveMarch2023}, as \ac{tether} is far less transparent about the nature and location of its reserves\cite{London2023StablecoinUSDT}.

We find an apparent flight-to-safety behavior: \ac{usdc}'s \ac{tvl} decreased $\sim$19.40\% relative to \ac{usdt}, while \ac{usdt}'s liquidity cost increased $\sim$241\%. Stablecoin-only pools lost liquidity providers as USDx-WETH/WBTC pools attracted more. These results are consistent with traditional financial distress literature about flight-to-safety behavior \cite{Lehnert2022Flight-to-safetyBehavior}, with investors rebalancing their portfolios towards seemingly
safer and more liquid assets \cite{Brunnermeier2009MarketLiquidity,Tobin1958LiquidityRisk} and the trend of decreasing liquidity during market turmoil \cite{Brunnermeier2009MarketLiquidity}. Additionally, we find that higher-fee pools tended to have considerably more providers (e.g., USDCWETH3000, WETHUSDT3000) than lower-fee ones (e.g., WBTCUSDC500, DAIUSDT500), but overall liquidity remained concentrated (Gini > 0.9), supporting findings in \cite{Lehar2023FragmentationExchanges}. Therefore, our results suggest that \ac{usdc}'s transparency and detailed disclosures led to swift reactions, while \ac{usdt}'s less frequent and opaque reporting provided a temporary buffer against immediate impacts.

%% file: sections/related_work.tex
\section{Related Work}

While existing literature has examined \ac{svb}'s collapse impact on global stock markets \cite{Pandey2023RepercussionsMarkets}, \ac{us} market sectors \cite{Yousaf2023ResponsesImplosion}, euro area banks \cite{Perdichizzi2023Non-significantFallout}, financial contagion in major economies \cite{Akhtaruzzaman2023DidContagion}, and cryptocurrency markets \cite{Galati2024SiliconStability}, our research uniquely focuses on the \ac{defi} sector and its spillover effects from \ac{tradfi} market shocks. The closest to our work is \cite{Galati2024SiliconStability}, who use a BEKK-GARCH model to analyze cryptocurrency price changes on \ac{cexs}. We instead examine liquidity dynamics within \ac{defi}, specifically \ac{usdc} and \ac{usdt} liquidity pools on Uniswap. Liquidity is crucial in financial markets, as its abundance or scarcity can determine the efficiency of price discovery, the speed at which new information is digested by the market, and even trigger catastrophic \blockquote{flash crashes} \cite{Brunnermeier2009MarketLiquidity,Chen2021LiquidityCrises,Amihud2005LiquidityPrices,Cornett2012LiquidityCrisis,Vayanos2012LiquidityCompetition,Filimonov2012QuantifyingCrashes,Kirilenko2017TheMarket}. While \ac{defi} markets are shallower and have lower volume than \ac{cexs}, they offer full transparency, and liquidity manipulation practices like order spoofing are much harder and riskier to implement than in unregulated \ac{cexs}, providing a market picture that is more likely to be genuine. Therefore, our study's significance lies in its analysis of \ac{defi} liquidity pool reactions to external shocks, which, to the best of our knowledge, is a previously unexplored area.

The underlying assumption that liquidity should react to shocks is well-rooted in traditional financial theory. Drechsler et al. \cite{Drechsler2020LiquidityVolatility} show that liquidity providers are negatively exposed to increases in volatility due to growing adverse selection risk: as such, a shock affecting the value of a stablecoin should map to a significant reduction of its available liquidity on the market. Chordia et al. \cite{Chordia2005AnLiquidity} find that volatility is informative in predicting liquidity shifts, while Amihud \cite{Amihud2002IlliquidityEffects} finds that expectations about liquidity affect valuations in stocks.

The effects of disclosure and transparency have also been widely studied in the context of traditional financial markets. The primary inspiration for our work is Holmstrom \cite{Holmstrom2019UnderstandingSystem}, who points out that the role of transparency in debt and monetary instruments is opposite to the one it has for equities. While in the latter higher transparency and disclosure are often associated with lower financing costs and higher valuations \cite{DIAMOND1991DisclosureCapital}, the former behave as \blockquote{no questions asked} assets, for which the only point when information is relevant is close to the maturity of the debt or the redemption of the monetary asset, impacting valuation exclusively in a negative or neutral fashion. Mario Draghi's \blockquote{whatever it takes} speech in July 2012 \cite{Draghi2012VerbatimDraghi} was a clear example of this theory at work, shrouding the deteriorating health of the Eurosystem behind a veil of opacity and thus saving it from what might have become a self-induced financial catastrophe for the European banking system.

Finally, we build on the growing literature on the microstructural properties of \ac{dexs}. In particular Lehar and Parlour \cite{Lehar2021DecentralizedMaker} compare \ac{amm}s with \ac{lob}-based exchanges and find market regimes under which \ac{amm}s are more convenient trading venues; Lehar et al. \cite{Lehar2023FragmentationExchanges} show that Uniswap v3 pools attract additional liquidity through market fragmentation; and \cite{Capponi2023TheLiquidity} highlight strategic liquidity provision practices that take advantage of the unique setting of public blockchains. Therefore, our findings contribute to the ongoing debate over the viability of \ac{defi} markets as complementary venues to their traditional counterparts.

%% file: sections/background.tex
\section{Background}

\subsection{Events} \label{subsec:events}

We analyze the market dynamics surrounding \ac{svb}'s collapse in March 2023, precipitated by the following events:

\begin{itemize}
    \item March 2022: \ac{fomc} starts increasing interest rates to combat inflation \cite{FEDInterestRates,FEDSpeechMarch2022}, affecting leveraged sectors and lending institutions \cite{FEDLeverageFinancialSectorMarch2023}.
    \item 8 March 2023: Silvergate Capital announces liquidation \cite{2023SilvergateBank,Lang2023Crypto-focusedReuters}.
    \item 9 March 2023: \ac{svb}'s stock falls more than 60\% at the stock market opening \cite{Hu2023SiliconReuters}.
    \item 10 March 2023: \ac{svb} experiences a bank run and regulatory takeover \cite{JoshuaFranklin2023SiliconRegulators}.
    \item 11 March 2023, 3:11 AM UTC: \ac{circle} reveals \$3.3 billion (8\% of its \$40 billion cash reserves \cite{USDCReserveMarch2023}) held at \ac{svb} \cite{Capoot2023StablecoinExposure}\footnote{https://twitter.com/circle/status/1634391505988206592\label{footnote:circleAnnouncement}}.
    \item 12 March 2023: Signature Bank closure by New York regulators \cite{Lang2023SignatureReuters}.
    \item 17 March 2023: \ac{svb}'s parent company files for chapter 11 bankruptcy \cite{KrollSVBChapter11}.
    \item 22 March 2023: the \ac{fed} raises rates to 4.75-5\% \cite{ColbySmith2023FederalTurmoil}.
    \item 26 March 2023: \ac{fcb} acquires \ac{svb} \cite{FCBBoughtSVBPressRelease2023}.
\end{itemize}

We define three analysis periods relative to \ac{circle}'s announcement\footref{footnote:circleAnnouncement}:

\begin{itemize}
    \item \textbf{Before}: 1 February 2023 -- 3:11 AM UTC, 11 March 2023
    \item \textbf{During}: 3:11 AM UTC, 11 March 2023 -- 17 March 2023
    \item \textbf{After}: 17 March 2023 -- 30 April 2023
\end{itemize}

\subsection{Liquidity and exchange mechanisms in \ac{defi}}

\subsubsection{Mechanism} \label{subsec:data_preprocessing}

A liquidity position ($L^2$) in Uniswap's V3 is defined by its \ac{amm}'s equation \cite{Adams2021UniswapCore}: 
\begin{equation} \label{eq:UniswapV3AMM}
    \underbrace{(X_{real} + \frac{L}{\sqrt{P_b}})}_{X_{virtual}}\overbrace{(Y_{real} + L\sqrt{P_a})}^{Y_{virtual}} = L^2
\end{equation}
Re-arranging \autoref{eq:UniswapV3AMM}, we can calculate the real amounts of token $X$ ($X_{real}$) and token $Y$ ($Y_{real}$) in tick $i$, identified by its price bounds $[P_{a}^{(i)}, P_{b}^{(i)}]$ when:

\begin{equation} \label{eq:UniswapV3LiquidityTicks}
    X_{real}, Y_{real} = 
    \begin{cases}
        L \frac{\sqrt{P_b^{(i)}}- \sqrt{P_a^{(i)}}}{\sqrt{P_a^{(i)} P_b^{(i)}}},  0 & \text{if } P < P_a^{(i)} \\
        0, L(\sqrt{P_b^{(i)}}-\sqrt{P_a^{(i)}}) & \text{if } P \geq P_b^{(i)} \\
        L \frac{\sqrt{P_b^{(i)}}- \sqrt{P}}{\sqrt{P} \times \sqrt{P_b^{(i)}}}, L(\sqrt{P}-\sqrt{P_a^{(i)}}) & \text{if } P_a^{(i)} \leq P < P_b^{(i)} 
    \end{cases}
\end{equation}

\subsubsection{Exchange}

Traditional \ac{lob} exchanges match orders based on price and time priority, while \ac{amm} exchanges, like Uniswap and others, use a constant product formula (\autoref{eq:UniswapV3AMM}) to determine prices from token ratios in liquidity pools \cite{Adams2020UniswapV2Core,Xu2023SoK:Protocols,Dahi2023AutomatedLedger}. Despite this difference, price effects in both systems are comparable, with a \ac{lob} market's midpoint analogous to an \ac{amm} pool's current price \cite{Lehar2023FragmentationExchanges}. Additionally, Uniswap has several versions, but for our analysis, we focus on Uniswap V3 due to its other advantages, some of which are analogous to \ac{lobs}:

\begin{itemize}
    \item Concentrated liquidity within price range $[P_{a}, P_{b}]$, unlike V1 and V2's $[0,\infty]$ distribution \cite{Adams2020UniswapV2Core}, which means that trades execute against liquidity within a specified price range $[P_{a}, P_{b}]$  \cite{Adams2021UniswapCore,Fan2021StrategicV3}, similar to a market maker's simultaneous sell and buy orders in a \ac{lob} \cite{Lehar2023FragmentationExchanges}.
    \item Higher trading volume and more responsive liquidity provision \cite{TopDEXsCoinMarketCap}.
    \item Multiple fee tiers (0.01\%, 0.05\%, 0.30\%, 1.00\%) for risk-reward adjustment \cite{Adams2021UniswapCore}.
\end{itemize}

\subsubsection{Liquidity Provision in Uniswap}

Liquidity is crucial in financial markets, enabling easy buying and selling of assets without significant price changes \cite{Amihud1986AssetSpread}. An \ac{amm} pool in Uniswap's \ac{dex} acts as the sole market maker, separating liquidity providers from traders \cite{Adams2021UniswapCore}. This structure may create a more level playing field for traders \cite{Angeris2020ImprovedMakers}. Additionally, Uniswap's transparency, by running on a blockchain, allows the identification of liquidity providers through \ac{nfts} representing their positions \cite{UniswapPositionsNFT} and also, enabling all participants to see available liquidity, potentially improving price discovery and reducing information asymmetry \cite{Adams2021UniswapCore,Xiong2023DemystifyingV3,Capponi2023TheLiquidity}. This transparency may encourage responsible market-making and better monitoring of manipulation \cite{Makarov2021NBERMARKET}, but could expose providers to front-running or targeted attacks \cite{Eskandari2020SoK:Blockchain}. For example, a small fraction ($\sim$0.3\%) of Uniswap V3 liquidity comes from Miner Extractable Value (MEV) bots executing Just-In-Time (JIT) liquidity attacks \cite{Wan2022Just-in-timeProtocol,Xiong2023DemystifyingV3,Capponi2023TheLiquidity}, which, however, we consider negligible for our study.

%% file: sections/methodology.tex
\section{Methodology}

\subsection{Data Collection}

We analyze ten key Uniswap liquidity pools (\autoref{tab:did_pools}) representing our control (\ac{usdt}) and treatment (\ac{usdc}) groups, selected for their consistently high \ac{tvl} and volume \cite{Lehar2021DecentralizedMaker,USDCTopLiquidityPools,USDTTopLiquidityPools}, at the time of this writing. Our selection criteria prioritize pairs with \ac{weth}\footnote{\ac{weth} has the same value as ETH, its underlying asset}, which typically have more liquidity on Uniswap than on \ac{cexs} \cite{Liao2022TheLiquidity}, as well as \ac{dai} and \ac{wbtc}\footnote{\ac{wbtc} has the same value as BTC, its underlying asset} pairs due to their high volume and \ac{tvl} on Uniswap \cite{TopPoolsUniswapV3}. We also include the \ac{usdc}/\ac{usdt} pair for direct comparison between treatment and control groups. To reconstruct liquidity pool states for the \textit{before}, \textit{during}, and \textit{after} periods (\autoref{subsec:events}), we:

\input{tables/LiquidityPools}

\begin{enumerate}
    \item Obtain latest positions from Uniswap V3's subgraph\footnote{https://github.com/Uniswap/v3-subgraph\label{footnote:UniswapV3SubgraphLink}} at current time $T_0$ (data collection start).
    \item Trace positions backward ($T_0, T_{-1}, T_{-2}, \ldots, T_{-n}$).
    \item Identify and record closed (burned) positions via unique \ac{nft} identifiers.
    \item Add burned positions back to the reconstructed pool state at relevant times ($T_{-n}$).
\end{enumerate} 

This process reliably reconstructs historical liquidity pool states based on the data available at Uniswap V3's subgraph\footref{footnote:UniswapV3SubgraphLink}.

\subsection{Liquidity analysis metrics}

\subsubsection{\acl{mci}}

To analyze the impact of events on liquidity costs, we take advantage of the similarities between Uniswap V3 pools' concentrated liquidity architecture and a more traditional \ac{lob} \cite{Lehar2023FragmentationExchanges}. We then adapt %
the \acl{mci} measure introduced by Cenesizoglu et al. \cite{Cenesizoglu2018Bid-Book} to a liquidity pool, in which the ask side cost of liquidity is defined as:
\begin{equation} \label{eq:MCI_A_LOB}
MCI_A = \frac{VWAPM_A}{Vlm_A}
\end{equation}
\begin{equation}
VWAPM_A = \ln \frac{\frac{Vlm_A}{\sum_{l=1}^{L} Q_{A,l}}}{0.5(P_{A,1} + P_{B,1})}
\end{equation}
\begin{equation}
Vlm_A = \sum_{l=1}^{L} P_{A,l}Q_{A,l}
\end{equation}

and for the bid side as:
\begin{equation} \label{eq:MCI_B_LOB}
MCI_B = \frac{-VWAPM_B}{Vlm_B}
\end{equation}

\ac{mci} measures the marginal cost of executing trades that consume a significant portion of available liquidity, considering its distribution across price levels \cite{Cenesizoglu2018Bid-Book}. This concept applies to both \ac{lob} and \ac{amm} exchanges, as it captures the ease of an asset's trading without causing significant price movements. By incorporating price and quantity information for \ac{usdc} and \ac{usdt} from their liquidity pools, we can measure transaction costs during \ac{svb}'s fallout. In Uniswap's \ac{amm} context:
\begin{itemize}
    \item Buy orders: Swap the paired token for \ac{usdc} or \ac{usdt} (e.g., \ac{weth} for \ac{usdc} in the \ac{usdc}/\ac{weth} pool).
    \item Sell orders: Swap \ac{usdc} or \ac{usdt} for the paired token.
\end{itemize}

We adapt the \ac{mci} formula (\autoref{eq:MCI_A_LOB} and \autoref{eq:MCI_B_LOB}) for Uniswap by considering liquidity at different price levels within $[P_{a}, P_{b}]$, as determined by active liquidity. To calculate $VWAP$, we simulate order execution with given sizes or tick spans. First, to allocate $Y$ and $X$ tokens into ticks at a given time, we aggregate liquidity ($L^2$) (\autoref{eq:UniswapV3AMM}) for all positions in tick $i$ at time $T_{0-n}$ using \autoref{eq:UniswapV3LiquidityTicks}. Then, for a sell or buy order consuming all liquidity in tick $i$, we calculate $\Delta X$ and $\Delta Y$ using \autoref{eq:UniswapV3Swapping}:

\begin{equation} \label{eq:UniswapV3Swapping}
    \Delta X, \Delta Y = 
    \begin{cases}
        X_{real},  \frac{L^2}{X_{virtual} - \Delta X} - Y_{virtual} & \text{for a sell order (swap token $Y$ for $X$)} \\
        \frac{L^2}{Y_{virtual} - \Delta Y} - X_{virtual}, Y_{real} & \text{For a buy order (swap token $X$ for $Y$)}
    \end{cases}
\end{equation}

By calculating the swap $\Delta X$ and $\Delta Y$ for a given order size consuming all the liquidity at a level or range of levels, the \ac{mci} formula estimates the cost of executing a buy or sell order given a specific liquidity level in the affected price range. This is analogous to how \ac{mci} is calculated for \ac{lob} exchanges, where the formula considers the available liquidity at different order book levels. We can then adapt the 
\ac{mci} of \cite{Cenesizoglu2018Bid-Book} to the DeFi case by recognizing that the Volume-Weighted Average Price ($VWAP$) and the Volume-Weighted Average Price scaled by the Mid-price ($VWAPM$), which in our case is the pre-transaction price $P$ on the pool, are:
\begin{equation}
VWAPM = \ln \frac{\frac{\sum_l \Delta X_l}{\sum_l \Delta Y_l}}{P}
\end{equation}
Finally, the \ac{mci} for buy and sell orders is calculated as:
\begin{equation}
MCI = (-1)^{B} \times \frac{VWAPM}{\sum_l \Delta X_l}
\end{equation}

where $B$ is 1 for sell (bid-side) orders and 0 for buy (ask-side) orders to calculate $MCI_B$ and $MCI_A$, respectively. Consistent with the literature \cite{Cenesizoglu2018Bid-Book}, we represent \ac{mci} in basis points per thousand $X$ units, which in our case is a stablecoin. Once we generate the $MCI_A$ and $MCI_B$, we can calculate the bid-ask imbalance \cite{Cenesizoglu2018Bid-Book}, denoted as $MCI_{IMB}$, by:

\begin{equation}
    MCI_{IMB} = \frac{MCI_A-MCI_B}{MCI_A+MCI_B}
\end{equation}

A positive imbalance implies that the marginal cost of swapping ask-side liquidity (buying) is higher than the cost of swapping bid-side liquidity (selling), and vice-versa. Finally, we also calculate the average $MCI_{\mu}$, denoted as:
\begin{equation}
    MCI_{\mu} = \frac{MCI_A + MCI_B}{2}
\end{equation}

which we use to quantify the average cost of liquidity regardless of the transaction side.

\subsection{Event study}

\input{tables/didResults}

\input{tables/DiDMCI}

We employ an event study methodology to assess the impact of key events (\autoref{subsec:events}) on liquidity costs, as well as the number of active liquidity providers and their liquidity concentration measured by a Gini coefficient. Using \ac{did}, we measure the significance of these changes, with \ac{usdc} pools (\autoref{tab:did_pools}) as the treatment group and the same-pair \ac{usdt} pools as the control group. This approach isolates the effect of \ac{svb}'s downfall on \ac{usdc}'s top liquidity pools, represented as:

\begin{equation}\label{eq:did}
    y_{i,t} = \beta + \beta_1 \cdot {\bf 1}_{t>\tau} + \beta_2 \cdot {\bf 1}_{i=\text{USDC}} + \beta_3 \cdot (
    {\bf 1}_{i=\text{USDC}} \times {\bf 1}_{t>\tau}
    ) + \epsilon_{i,t}
\end{equation}

where
$\tau$ is the treatment date,
${\bf 1}_A$ is $1$ if $A$ is true and $0$ otherwise, and $y$ can be \ac{tvl} in \ac{usd} or $\ac{mci}_\mu$ measured on the first 1, 5, 10, 15 or 20 liquidity pool ticks around the active tick. From \autoref{eq:did}, we care about the statistical significance of $\beta_3$ and the interaction between ${\bf 1}_{i=\text{USDC}}$ and ${\bf 1}_{t>\tau}$. Finally, the ratio $\frac{\beta_3}{\beta_2}$ quantifies the net effect that the events had on \ac{usdc} pools and not on \ac{usdt}.

%% file: tables/LiquidityPools.tex
\begin{table}[tb]
    \centering
    \scriptsize %
    \caption{Liquidity Pools for \ac{did} analysis with trading fees}
    \begin{tabular}{ll}
    \toprule
        \textbf{Control (\ac{usdt})} & \textbf{Treatment (\ac{usdc})} \\
        \midrule
        WETH/USDT (Fee: 0.01)    & USDC/WETH (Fee: 0.01)     \\
        WETH/USDT (Fee: 0.05)    & USDC/WETH (Fee: 0.05)     \\
        WETH/USDT (Fee: 0.3)     & USDC/WETH (Fee: 0.3)      \\
        WBTC/USDT (Fee: 0.3)    & WBTC/USDC (Fee: 0.05)    \\
        DAI/USDT (Fee: 0.05)    & DAI/USDC (Fee: 0.05)     \\
        \midrule
        \multicolumn{2}{c}{USDC/USDT (Fee: 0.01)} \\
    \bottomrule
    \end{tabular}
    \label{tab:did_pools}
\end{table}

%% file: tables/DiDResults.tex
\begin{table}[bt]
\scriptsize
\centering
\caption{Differences-in-Differences Estimation Results}
\makebox[\textwidth]{
\begin{tabular}{lcc}
\toprule
& \acl{tvl} in \acs{usd} \\
\midrule
Treatment ($\beta_1$) & $0.067217^{***}$ \\
& $(0.017)$ \\
Group ($\beta_2$) & $1.668278^{***}$ \\
& $(0.018)$ \\
Treatment interaction ($\beta_3$) & $-0.323698^{***}$ \\
& $(0.024)$ \\
\addlinespace
Relative Effect $\frac{\beta_3}{\beta_2}$ & -0.1940 \\
Observations & 178 \\
\bottomrule
\end{tabular}
}
\begin{tablenotes}
\scriptsize
\centering
\item Standard errors are in parentheses. $^{*}p<0.05$, $^{**}p<0.01$, $^{***}p<0.001$
\end{tablenotes}
\label{tab:did_model_results}
\end{table}

%% file: tables/DiDMCI.tex
\begin{table}[hbt!]
\centering
\scriptsize
\caption{Differences-in-Differences Estimation Results using \ac{mci} for liquidity pool levels 1, 5, 10, 15, 20}

\begin{tabular}{lccccc}
\toprule
& 1 & 5 & 10 & 15 & 20 \\
\midrule
Treatment ($\beta_1$) & $-0.0138^{***}$ & $-0.0137^{***}$ & $-0.0137^{***}$ & $-0.0137^{***}$ & $-0.0136^{***}$ \\
& $(0.001)$ & $(0.003)$ & $(0.001)$ & $(0.001)$ & $(0.001)$ \\
Group ($\beta_2$) & $0.0035^{*}$ & $0.0017$ & $0.0015$ & $0.0015$ & $0.0016$ \\
& $(0.001)$ & $(0.004)$ & $(0.001)$ & $(0.001)$ & $(0.001)$ \\
Treatment Interaction ($\beta_3$) & $0.0166^{***}$ & $0.0207^{***}$ & $0.0172^{***}$ & $0.0162^{***}$ & $0.0155^{***}$ \\
& $(0.002)$ & $(0.004)$ & $(0.002)$ & $(0.002)$ & $(0.002)$ \\
\addlinespace
Relative Effect $\frac{\beta_3}{\beta_2}$ & 4.8012 & 11.9550 &  11.6857 &  10.7729 &  9.5653 \\
Observations & $1442$ & $1442$ & $1442$ & $1442$ & $1442$ \\
\bottomrule
\end{tabular}

\begin{tablenotes}
\scriptsize
\centering
\item Standard errors are in parentheses. $^{*}p<0.05$, $^{**}p<0.01$, $^{***}p<0.001$
\end{tablenotes}
\label{tab:did_model_mci_mean}
\end{table}

%% file: sections/results.tex
\section{Results}

\subsection{\acl{did}} \label{subsec:did_results}

The statistically significant negative treatment interaction coefficient ($\beta_3$) for \acl{tvl} in \ac{usd} reveals the \ac{svb} collapse's substantial impact on the treated group (\ac{usdc}) compared to the control group (\ac{usdt}). The relative effect, calculated as $\frac{\beta_3}{\beta_2}$, weakens the advantage in \ac{tvl} of \ac{usdc} relative to \ac{usdt} by a $\sim$19.40\% post-event (see \autoref{tab:did_model_results}). This aligns with \cite{Ahmed2024PublicRuns,Galati2024SiliconStability}, demonstrating that stablecoins with perceived stronger ties to traditional banking are more susceptible to financial stress spillovers.

\subsection{\acl{mci}} \label{subsec:mci_results}

\begin{figure*}[bt!]
\centering
    \input{figures/MCI_levels_values}
    \hfill
    \input{figures/MCI_imbalance_levels}
    \caption{Median daily $MCI_{\mu}$ and $MCI_{IMB}$ for \ac{usdc} and \ac{usdt}. Shaded area: interquartile range (75th to 25th percentile). Lines mark events: Silvergate Bank's liquidation (teal), \ac{svb} stock crash (gray), Circle's tweet (tan), \ac{svb} bankruptcy (red), \ac{fed} rate hike (golden), \ac{fcb} buys \ac{svb} (green).}
    \label{fig:MCI_results}
\end{figure*}
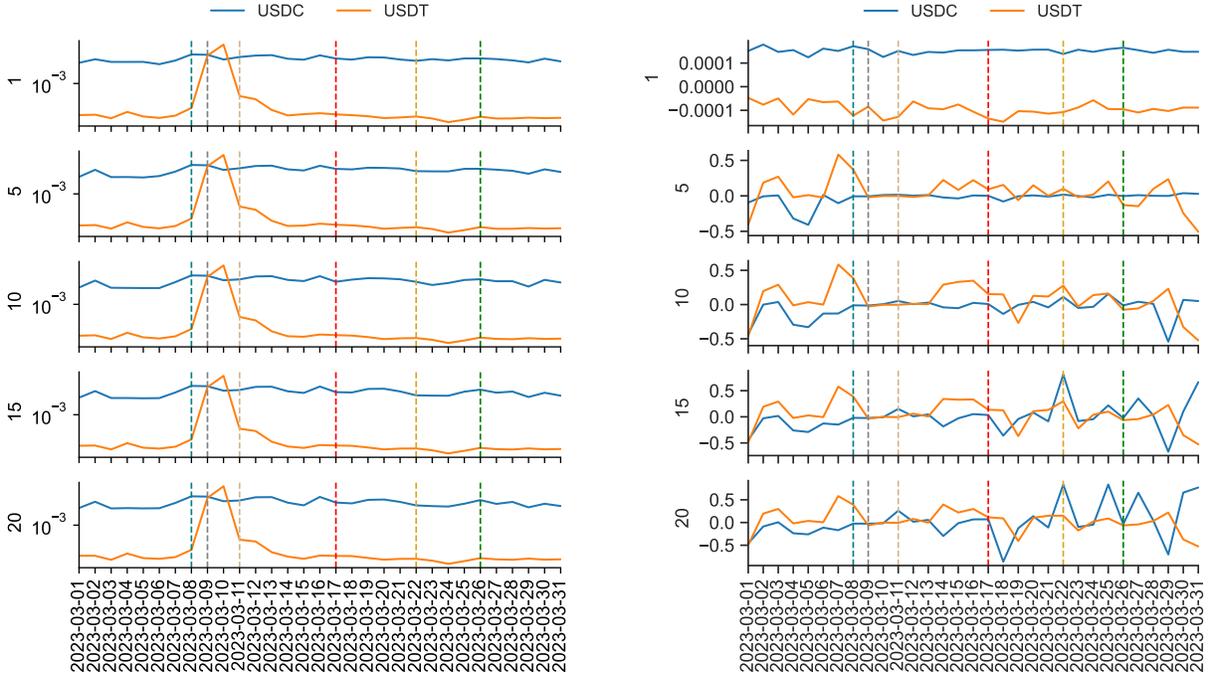

The \ac{did} estimation results for $\ac{mci}_\mu$ across liquidity pool levels 1, 5, 10, 15, and 20 (\autoref{tab:did_model_mci_mean}) reveal the differential impact of the \ac{svb} collapse on \ac{usdc} and \ac{usdt}, with \ac{usdc} experiencing a more significant increase in marginal trading costs, especially at deeper pool levels. The negative and statistically significant treatment coefficient ($\beta_{1}$) across all levels indicates lower $\ac{mci}_\mu$ values for \ac{usdc} (treated group) compared to \ac{usdt} (control group) during the event period. However, the positive and statistically significant treatment interaction coefficient ($\beta_{3}$) shows that the gap in $\ac{mci}_\mu$ values between \ac{usdc} and \ac{usdt} widened during this period, implying a more substantial increase in marginal trading costs for \ac{usdc}. The relative effect, calculated as $\frac{\beta_3}{\beta_2}$, increases from 4.80 at level 1 to 11.96 at level 5, indicating that the difference in marginal trading costs between \ac{usdc} and \ac{usdt} became more pronounced at deeper liquidity pool levels.

\subsection{Gini Coefficient}

\begin{figure*}[hbt!]
\centering    
    \input{figures/USDC_Gini}
    \hfill
    \input{figures/USDC_liquidity_providers}
    \hfill
    \input{figures/USDT_Gini}
    \hfill
    \input{figures/USDT_liquidity_providers}  
    \caption{Gini coefficient and total liquidity providers for \ac{usdc} and \ac{usdt}. The lines across the plots represent different events.}
    \label{fig:Gini_results}
\end{figure*}

The Gini Coefficient remained relatively high (above 0.9) for most \ac{usdc} and \ac{usdt} trading pairs throughout March 2023 (see \autoref{fig:USDCGiniCoeff} and \autoref{fig:USDTGiniCoeff}). This suggests a concentrated liquidity provision, with a small number of liquidity providers contributing a significant proportion of the total liquidity. However, some trading pairs, such as WBTCUSDC500, showed lower Gini Coefficients, indicating a more even distribution of liquidity among providers (see \autoref{fig:USDCGiniCoeff} and \autoref{fig:USDTGiniCoeff}) while other trading pairs like USDCUSDT100 had more activity regarding liquidity providers adding or removing liquidity. We analyze these results further on \autoref{subsubsec:liquidity_concentration}

%% file: figures/MCI_levels_values.tex
\begin{subfigure}[tb]{0.48\textwidth}
   \centering
     \centering
     \includegraphics[width=\linewidth]{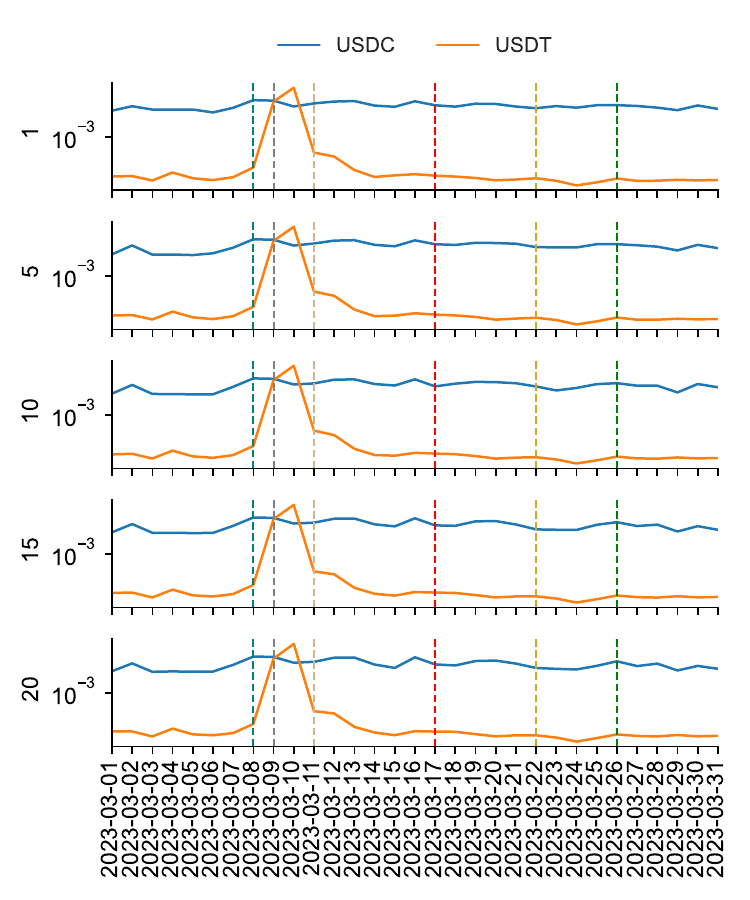}
       \caption{Median daily $MCI_{\mu}$ for \ac{amm}'s liquidity levels of 1, 5, 10, 15, and 20}
     \label{fig:MCImedianLiquidityLevelsValues}
 \end{subfigure}

%% file: figures/MCI_imbalance_levels.tex
\begin{subfigure}[tb]{0.48\textwidth}
   \centering
     \centering
     \includegraphics[width=\linewidth]{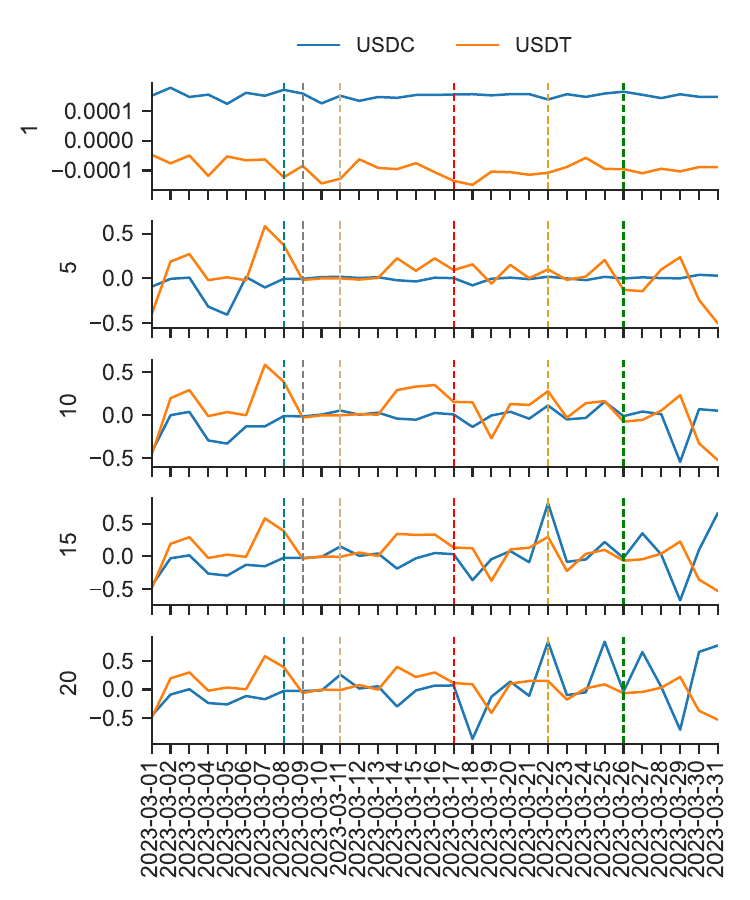}
    \caption{Median daily $MCI_{IMB}$ for \ac{amm}'s liquidity levels of 1, 5, 10, 15, and 20}
     \label{fig:MCIimbalanceLevels}
 \end{subfigure}

%% file: figures/USDC_Gini.tex
\begin{subfigure}[tb]{0.47\textwidth}
   \centering
     \centering
     \includegraphics[width=\linewidth]{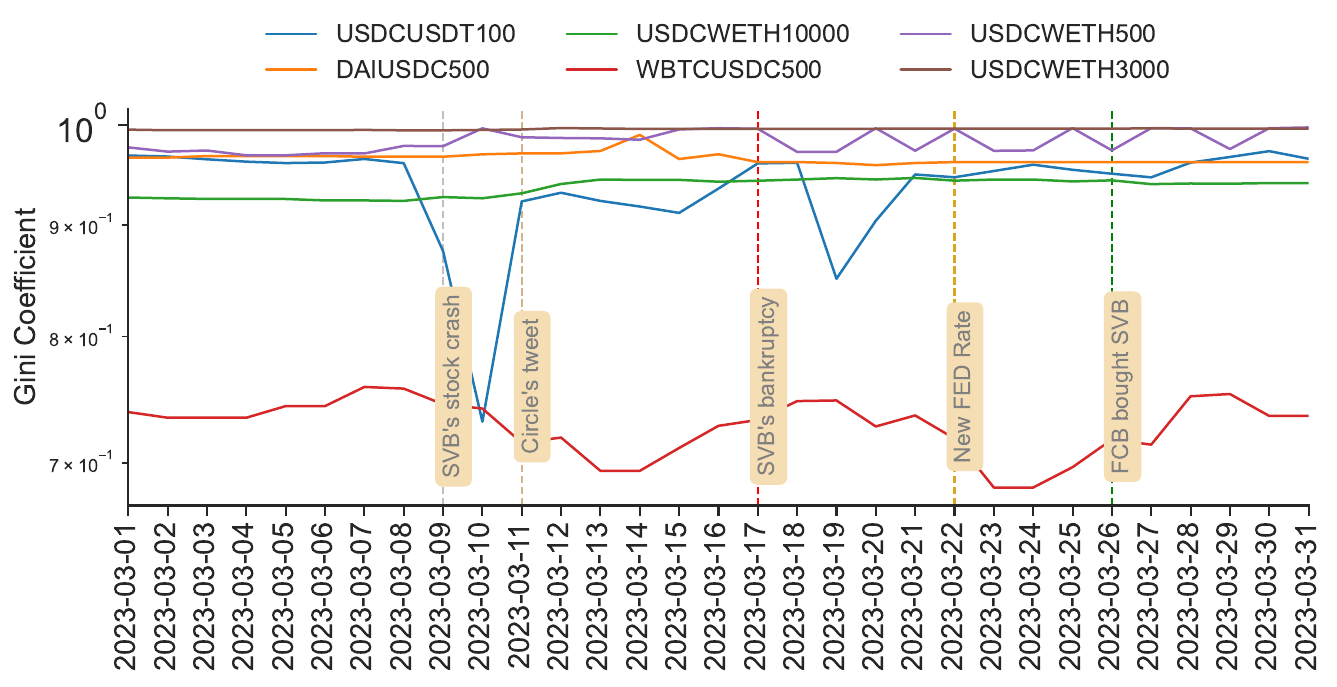}
     \caption{Median daily Gini Coefficient for \ac{usdc}'s liquidity pools (\autoref{tab:did_pools})}
     \label{fig:USDCGiniCoeff}
 \end{subfigure}

%% file: figures/USDC_liquidity_providers.tex
\begin{subfigure}[tb]{0.47\textwidth}
   \centering
     \centering
     \includegraphics[width=\linewidth]{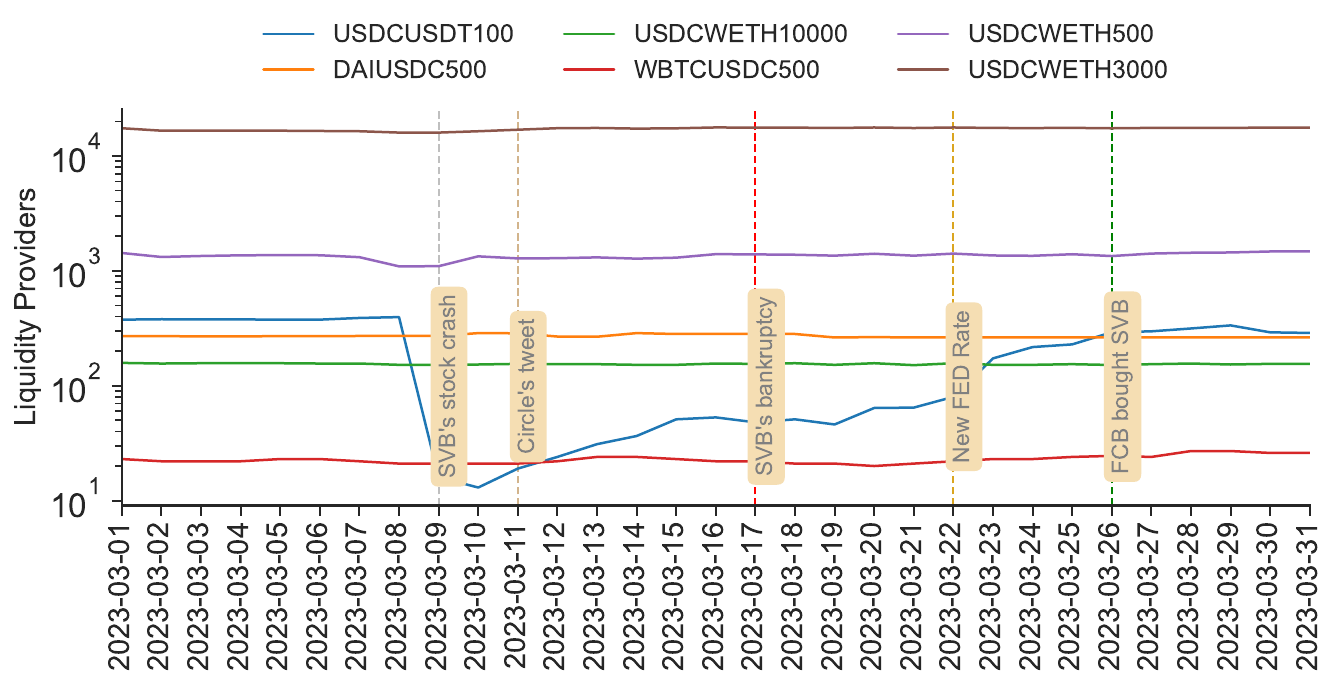}
     \caption{Median daily total liquidity providers for \ac{usdc}'s liquidity pools (\autoref{tab:did_pools})}
     \label{fig:USDCLiquidityProviders}
 \end{subfigure}

%% file: figures/USDT_Gini.tex
\begin{subfigure}[tb]{0.47\textwidth}
   \centering
     \centering
     \includegraphics[width=\linewidth]{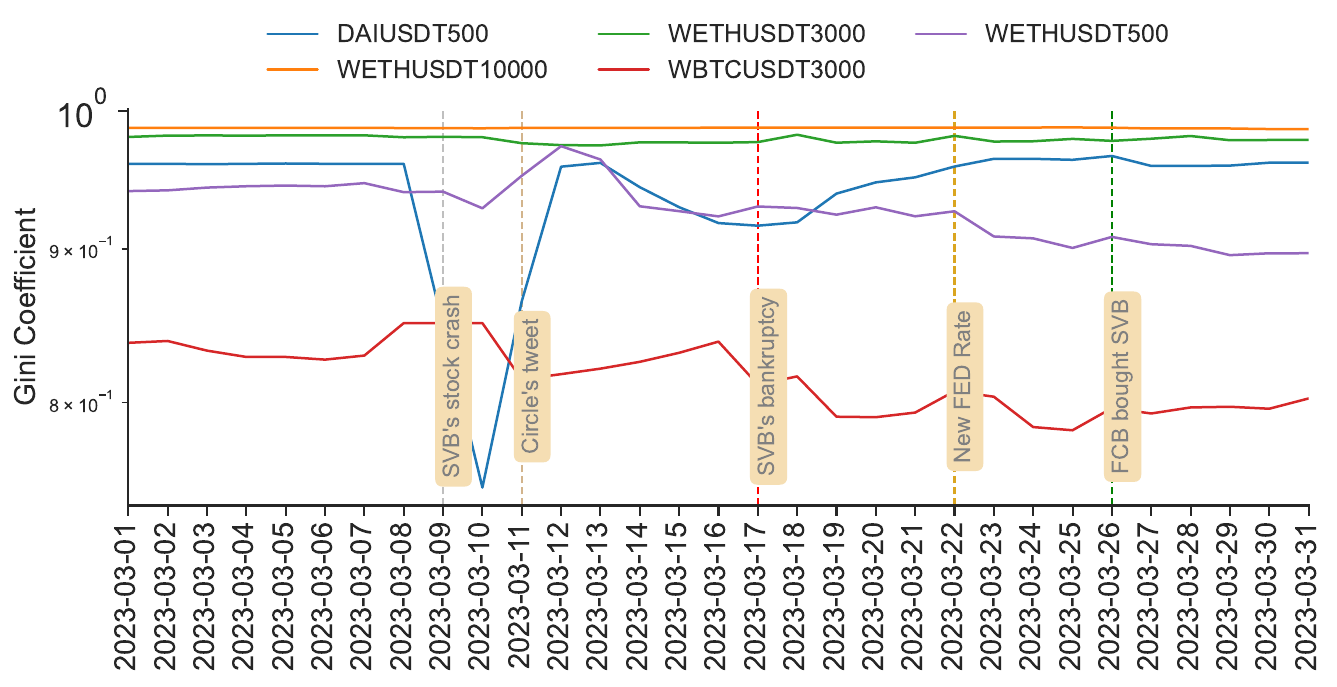}
     \caption{Median daily Gini Coefficient for \ac{usdt}'s liquidity pools (\autoref{tab:did_pools})}
     \label{fig:USDTGiniCoeff}
 \end{subfigure}

%% file: figures/USDT_liquidity_providers.tex
\begin{subfigure}[tb]{0.47\textwidth}
   \centering
     \centering
     \includegraphics[width=\linewidth]{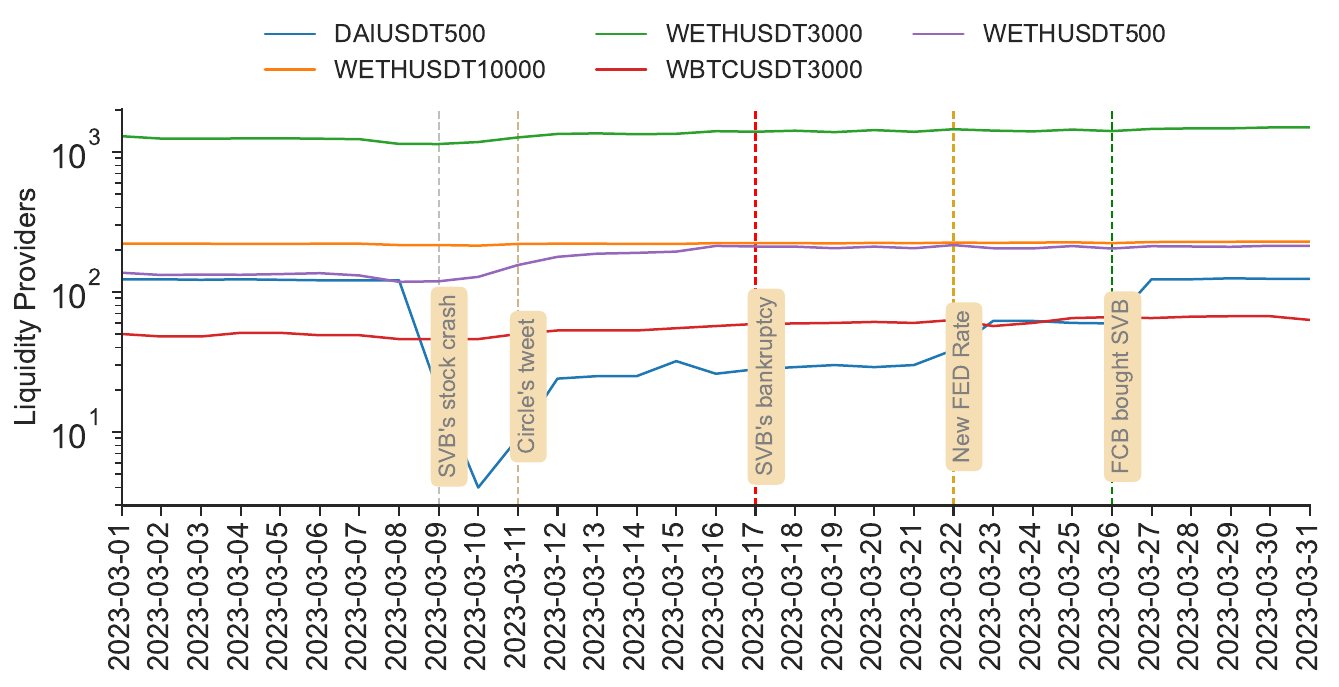}
     \caption{Median daily total liquidity providers for \ac{usdt}'s liquidity pools (\autoref{tab:did_pools})}
     \label{fig:USDTLiquidityProviders}
 \end{subfigure}

%% file: sections/discussion.tex
\section{Discussion} \label{sec:discussion}

\subsection{Liquidity analysis}\label{subsec:liquidity_analysis}

\subsubsection{\acl{mci}} \label{subsubsec:mci_analysis}

Our results in \autoref{subsec:mci_results} align with the liquidity preference theory \cite{Tobin1958LiquidityRisk}, which asserts that investors shift towards safer, more liquid assets during market stress \cite{Brunnermeier2009MarketLiquidity}. As \ac{usdc}'s exposure to \ac{svb} became known, investors perceived \ac{usdt} as a safer alternative, evidenced by changes in $\ac{mci}_{\mu}$ and $\ac{mci}_{IMB}$ between 8-18 March 2023. Following Silvergate Bank's liquidation announcement on 8 March, \ac{usdt}'s daily median $\ac{mci}_{\mu}$ at levels 1 and 20 increased by $\sim$241\%, indicating heightened buying pressure. This trend intensified on 9 March with \ac{svb}'s stock crash \cite{Hu2023SiliconReuters} but reversed on 11 March when Circle disclosed its \ac{svb} reserves\footnote[1]{https://twitter.com/circle/status/1634391505988206592}.

On 17 March, when \ac{svb} declared bankruptcy \cite{KrollSVBChapter11}, \ac{usdc}'s selling pressure increased significantly, with its $\ac{mci}_{IMB}$ at 20 levels dropping by $\sim$186\%. Conversely, \ac{usdt}'s positive $\ac{mci}_{IMB}$ suggested a buying preference, aligning with \cite{Galati2024SiliconStability}'s findings on stablecoins' vulnerability to financial stress. This flight-to-safety behavior increased liquidity demand for \ac{usdt}, despite its lack of transparency regarding cash reserves \cite{USDTReserveMarch2023,London2023StablecoinUSDT}. By 20 March, \ac{usdc}'s \ac{mci} imbalance (levels 10-20) matched \ac{usdt}'s, indicating ongoing market uncertainty. The \ac{fed}'s interest rate hike announcement on 22 March \cite{ColbySmith2023FederalTurmoil,Rushe2023USGuardian,Siegel2023FederalPost}, which seemingly led to increased buying pressure for \ac{usdc}. Then, despite \ac{fcb}'s announcement to buy \ac{svb} \cite{FCBBoughtSVBPressRelease2023} seemed to have improved market sentiment (\autoref{subsec:monetary_policy_and_stability}), \ac{mci} imbalances did not return to pre-event levels, suggesting a persistent impact on market sentiment and a continued preference for \ac{usdt} over \ac{usdc} for the weeks that followed.

\subsubsection{Liquidity Concentration} \label{subsubsec:liquidity_concentration}

\ac{usdt} pairs experienced less pronounced changes because of a possible perception of more stability. The \ac{svb} crisis in March 2023 significantly impacted liquidity distribution in \ac{usdc} and \ac{usdt} trading pairs. Following \ac{svb}'s stock crash on 9 March \cite{Hu2023SiliconReuters} and Circle's announcement on 11 March\footnote[1]{https://twitter.com/circle/status/1634391505988206592}, the Gini Coefficient for \ac{usdc} pairs decreased (\autoref{fig:USDCGiniCoeff}). For instance, USDCUSDT100's Gini Coefficient dropped from 0.9609 to 0.8756 between 8-9 March, while liquidity providers plummeted from $\sim$395 to $\sim$16, a $\sim$95.95\% decrease (\autoref{fig:USDCLiquidityProviders}).

This aligns with the tendency for liquidity to decrease during market stress \cite{Brunnermeier2009MarketLiquidity}. On the other hand, Uniswap V3's design allows liquidity providers to adjust positions during market downturns quickly \cite{Heimbach2022RisksProviders,Heimbach2022ExploringDistress}, potentially explaining the rapid withdrawal of liquidity. Additionally, \ac{usdt} pairs experienced less pronounced changes in the number of liquidity providers, except for DAIUSDT500, which dropped from $\sim$121 to $\sim$19 between 8-9 March and matched the trend of USDCUSDT100. Curiously, WETHUSDT500 and WBTCUSDT3000 followed a decreasing trend in their Gini Coefficient starting on 8 March as more liquidity providers seemed to migrate to these pools from pools that are of only stablecoin pairs like DAIUSDT500 and USDCUSDT100. In this context, ETH and BTC may have been perceived as more stable than stablecoins directly affected (\ac{usdc}) or with uncertain exposure (\ac{usdt}) to \ac{svb}  crisis. Besides, the diversification benefits of holding cryptocurrency assets alongside stablecoins may have motivated liquidity providers to rebalance their portfolios, consistent with modern portfolio theory \cite{Markowitz2009PortfolioSelection}. 

\ac{svb}'s bankruptcy declaration on 17 March \cite{KrollSVBChapter11} led to another decrease in liquidity providers, particularly in the USDCUSDT100 pool (\autoref{fig:USDCLiquidityProviders}), likely due to heightened market uncertainty. The \ac{fed}'s interest rate hike announcement on 22 March \cite{ColbySmith2023FederalTurmoil} saw USDCUSDT100's Gini Coefficient at 0.9467 with 80 liquidity providers (\autoref{fig:USDCGiniCoeff}, \autoref{fig:USDTGiniCoeff}). \ac{fcb}'s acquisition of \ac{svb} on 26 March \cite{FCBBoughtSVBPressRelease2023} had a stabilizing effect, with USDCUSDT100's liquidity providers surging from 80 to 292, a 265\% increase (\autoref{fig:USDCLiquidityProviders}, \autoref{fig:USDTLiquidityProviders}). The less volatile Gini Coefficients and the return of liquidity providers by late March 2023 indicate restored market confidence following \ac{fcb}'s intervention in the \ac{svb} collapse.

\subsection{Market dynamics}\label{subsec:market_dynamics}

\input{figures/totalValueLockedUSD}

The banking crisis of March 2023, exemplified by \ac{svb} stock crash on March 9 following its announcement of a \$1.8 billion loss \cite{Hu2023SiliconReuters}, had significant spillover effects on the \ac{defi} ecosystem (see \autoref{subsec:did_results}). \ac{circle}, the issuer of \ac{usdc}, had disclosed in its January and February 2023 reserve reports that it held cash reserves at \ac{svb}, Silvergate Bank (which voluntarily liquidated on March 8), and Signature Bank (which closed on March 12) \cite{USDCReserveJanuary2023,USDCReserveFebruary2023,2023SilvergateBank,Lang2023SignatureReuters}. These institutions were among the seven banks managing \ac{usdc}'s cash reserves \cite{USDCReserveFebruary2023} in March 2023.

Although \ac{circle} did not reveal the exact distribution of its reserves across these banks\cite{USDCReserveJanuary2023,USDCReserveFebruary2023}, as the banking crisis escalated\cite{Galati2024SiliconStability}, informed investors began withdrawing \ac{usdc} from \ac{defi} liquidity pools (see \autoref{fig:tvlUSD}), such as those on Uniswap, and exchanging it for other stablecoins. Initially, Silvergate Bank reported a \$1 Billion loss on 1 March 2023 and raised concerns about its ability to continue operating\cite{2023SilvergateBank}, which culminated with Silvergate Bank's voluntary liquidation on 8 March\cite{2023SilvergateBank}. This news alone caused an 11.25\% drop in \ac{usdc}'s \acl{tvl} in \ac{usd} on Uniswap. Then, on March 9, when \ac{svb}'s stock fell more than 60\% at the stock market opening \cite{Hu2023SiliconReuters} until \ac{circle} publicly acknowledged its exposure to \ac{svb} on 11 March\cite{Capoot2023StablecoinExposure}, \ac{usdc}'s \acl{tvl} in \ac{usd} in Uniswap fell by 6.95\% (\autoref{fig:tvlUSD}). These investors' behaviors during Silvergate Bank and \ac{svb}'s events highlight the existence of information asymmetry in the market, with some well-informed investors reacting more quickly to market developments\cite{Akerlof1970TheMechanism}.

The flight-to-safety or flight-to-quality behavior highlights the strong connections between traditional banking and \ac{defi}. This interconnection is a primary driver of the observed spillover effects and market reactions within \ac{defi} amid the banking turmoil in March 2023. The flight-to-safety behavior observed during this period is consistent with the literature on investor behavior during times of financial stress\cite{Lehnert2022Flight-to-safetyBehavior}, in which investors tend to rebalance their portfolios towards \textit{seemingly} safer and more liquid assets\cite{Brunnermeier2009MarketLiquidity,Tobin1958LiquidityRisk}. The significant drop in \ac{usdc} liquidity pool's \ac{tvl} in \ac{usd} (see \autoref{fig:tvlUSD}) and the increased demand for other stablecoins\cite{Galati2024SiliconStability} (see \autoref{subsec:mci_results}) demonstrates this phenomenon in the context of \ac{defi}.

\subsection{Monetary Policy and Market Stabilization Measures} \label{subsec:monetary_policy_and_stability}

Major central banks, including the \ac{fed}, European Central Bank, and Bank of England, raised interest rates between 2022-2023 to combat inflation \cite{FEDInterestRates,ECBInterestRates,BankEnglandInterestRates}, pressuring financial institutions \cite{Jiang2023MonetaryRuns}. This particularly affected \ac{svb}, which had benefited from previous low-rate policies \cite{Harari2024InterestIndicators}. The rate hikes devalued \ac{svb}'s bonds, contributing to its March 2023 bank run vulnerability \cite{Jiang2023MonetaryRuns}. Despite ongoing banking turmoil \cite{JoeRennison2023StocksTimes}, the \ac{fed} raised rates to 4.75-5\% on 22 March 2023 \cite{ColbySmith2023FederalTurmoil}. This decision negatively impacted bank stocks \cite{JoeRennison2023StocksTimes,DaniRomero2023Stock0.25} and seemed to cause an 8.53\% drop in \ac{usdc} liquidity pools' \acs{tvl} in \ac{usd} (\autoref{fig:tvlUSD}). Stability returned after \ac{fcb}'s announcement to acquire \ac{svb} on 26 March 2023 \cite{FCBBoughtSVBPressRelease2023}, facilitated by the \ac{fdic}'s provision of a contingent liquidity credit line \cite{FCBBoughtSVBPressRelease2023}. This action restored banking confidence and reduced uncertainty about \ac{circle}'s reserves. The ensuing recovery in \ac{usdc}'s liquidity pools (\autoref{fig:tvlUSD}) reflects the stabilizing effect of decisive interventions, echoing observations from the 2008 crisis \cite{GortonFEDPanicPrevention2013} and Ben Bernanke's Great Depression research on bank failures' role in deepening economic downturns \cite{Bernanke1983NonmonetaryDepression}.

\subsection{Reserve assets composition of \ac{usdc} and \ac{usdt}} \label{subsec:assets_usdc_usdt}

The March 2023 banking turmoil highlighted the trade-offs between liquidity and reserve portfolio management for \ac{usdc} and \ac{usdt}. \ac{circle}'s liquidity-focused approach for \ac{usdc}, with high cash reserves (25.39\% in February\cite{USDCReserveFebruary2023}, 12.47\% in March\cite{USDCReserveMarch2023}, and 1.32\% in April 2023\cite{USDCReserveApril2023}), prioritized meeting potential redemption demands. This aligns with asset-liability management principles \cite{BaselCommitteeonBankingSupervision2013BaselTools} but exposed \ac{usdc} to greater risk when three of its seven deposit-holding banks failed, particularly \ac{svb}\footnote[1]{https://twitter.com/circle/status/1634391505988206592\label{refnote1}}. Contrarily, \ac{tether}'s more diversified \ac{usdt} reserve portfolio, including less liquid assets (e.g., precious metals \cite{Dinh2022EconomicMarkets,Raza2023ForecastingApproach}) and more volatile assets (e.g., Bitcoins \cite{Bakas2022WhatMarket,Kuiper2022ResearchVolatile}), provided a buffer against the turmoil with \ac{usdt}'s March 2023 reserves show only 0.59\% (\$0.48 billion) in cash out of its \$81.83 billion portfolio\cite{USDTReserveMarch2023}, demonstrating the potential stability benefits of diversification\cite{Markowitz2009PortfolioSelection}. However, \ac{tether}'s approach carries risks. In a stablecoin run scenario \cite{Ahmed2024PublicRuns}, highly volatile assets could redeemed at a lower initial value and less liquid assets could be challenging to convert without significant losses. This mismatch between \ac{usdt}'s liabilities (stablecoins issued) and illiquid reserve assets creates a maturity transformation risk, a key vulnerability in traditional banking that can fuel runs \cite{Bologna2000WorkingPolicy,Diamond2001LiquidityBanking}. The contrasting impacts of the banking turmoil on \ac{usdc} and \ac{usdt} underscore the complex balance between maintaining liquidity for redemptions and reducing portfolio risks through diversification.

\subsection{Transparency in reporting}

The disparate impacts on \ac{usdc} and \ac{usdt} of the March 2023 banking crisis may originate from differences in transparency and reserve disclosure frequencies. While transparency typically improves cryptocurrencies' \ac{icos} success \cite{Howell2020InitialSales}, stablecoins present a contradiction. \ac{circle}'s monthly \ac{usdc} reports \cite{Schroeer2023AnalyticalUSDC} contrast with \ac{tether}'s semi-annual \ac{usdt} reports \cite{London2023StablecoinUSDT}, allowing \ac{usdc} holders to respond more swiftly to perceived risks. \cite{Ahmed2024PublicRuns} show that greater reserve transparency can increase run risk under pessimistic expectations and low conversion costs. Our \ac{mci} liquidity analysis (\autoref{subsec:mci_results}) corroborates this during the March 2023 crisis. \ac{usdc}'s transparency about its \ac{svb} holdings could have likely reduced confidence, which was exacerbated by low \ac{defi} transaction costs and the inability to halt trading on \ac{dexs} like in \ac{tradfi} for particular assets or securities to prevent further loss of value \cite{SECTradingSuspension}. This aligns with theories on how information asymmetries amplify investor coordination failures \cite{Angeletos2006CrisesVolatility}. Contrarily, \ac{tether}'s March 2023 report lacked details on cash reserve deposits \cite{USDTReserveMarch2023,London2023StablecoinUSDT}, such as the banks holding them, seemingly obscuring its exposure. Despite \ac{usdt}'s opacity, traders sought more liquidity in its pools, showing a willingness to pay a premium to switch from \ac{usdc} to \ac{usdt} (\autoref{subsec:liquidity_analysis}, \autoref{subsec:market_dynamics}).

%% file: figures/totalValueLockedUSD.tex
\begin{figure}[htb]
   \centering
     \centering
     \includegraphics[width=0.65\textwidth]{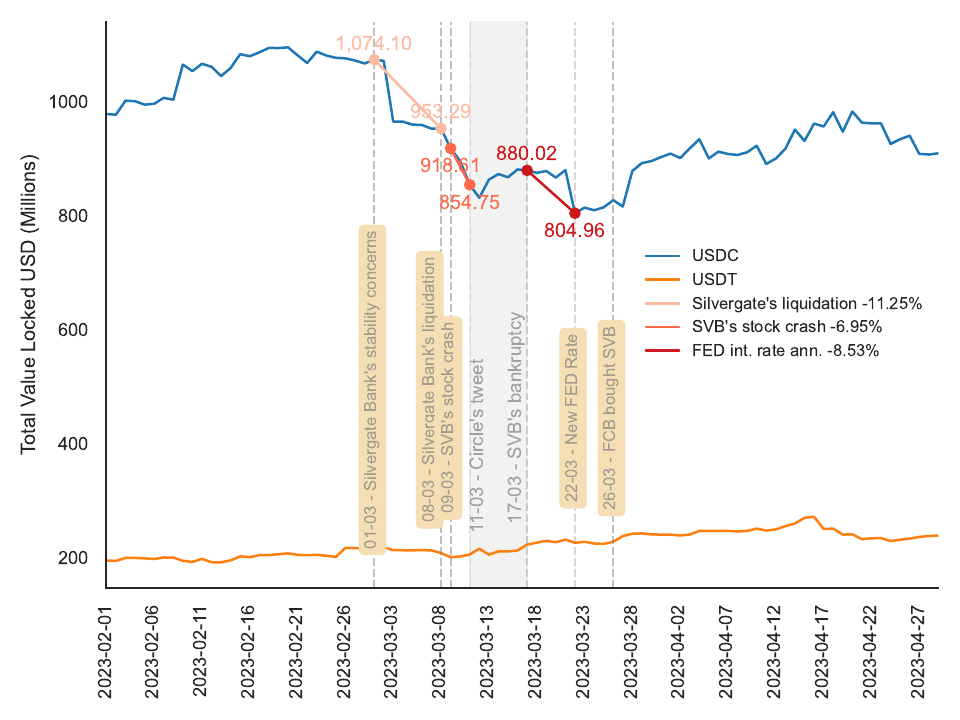}
     \caption{\acl{tvl} in \ac{usd} before and after different events}
     \label{fig:tvlUSD}
 \end{figure}

%% file: sections/limitations.tex
 \section{Limitations}

Our analysis focuses on \ac{usdc} and \ac{usdt}, including \ac{dai} as a trading pair, but does not capture interactions with other stablecoins like \ac{tusd}, etc. The study is limited to some of the top ten liquidity pools for \ac{usdc} and \ac{usdt} (\autoref{tab:did_pools}) at the time of writing. The dataset's hourly and daily frequency may not reflect sudden changes observable at more granular levels, as aggregation smooths out rapid fluctuations compared to higher-frequency data (e.g., minute-by-minute).

%% file: sections/conclusion.tex
\section{Conclusion}

The \ac{svb} collapse significantly impacted the \ac{defi} ecosystem. During this event, we observed an apparent flight-to-safety behavior: \ac{usdc}'s \ac{tvl} decreased $\sim$19.40\% relative to \ac{usdt}, \ac{usdt}'s liquidity cost increased by $\sim$241\%, while Stablecoin-only pools lost liquidity providers as USDx-ETH/BTC pools gained. These results align with traditional financial stress literature \cite{Lehnert2022Flight-to-safetyBehavior,Brunnermeier2009MarketLiquidity,Tobin1958LiquidityRisk}. 

Our findings revealed a transparency contradiction in stablecoins, suggesting that \ac{usdc}'s transparency and high-frequency reporting led to swift and abrupt market reactions, while \ac{usdt}'s less frequent and opaque disclosures provided a temporary buffer. The results would suggest the need to re-evaluate disclosure policies for stablecoins that have no safety net (e.g. insurance or lenders of last resort) regarding their reserves liquidity, as well as a stronger focus on robust reserve management to reduce liquidity risks.%